\documentclass[
aip,
jap,
numerical,
reprint,
twocolumn,
showpacs,
amsmath,
amssymb]{revtex4-1}
\usepackage{graphicx}
\usepackage{dcolumn}
\usepackage{bm}
\usepackage{textcomp}
\usepackage{revsymb}
\usepackage{amssymb}
\usepackage{ifsym}
\usepackage{wasysym}
\usepackage{dcolumn}
\usepackage{txfonts}

\begin{document}


\title{Pt silicide/poly-Si Schottky diodes as temperature sensors for bolometers}

\author{V. A. Yuryev}
\email{vyuryev@kapella.gpi.ru} 
\homepage[home page: ]{http://www.gpi.ru/eng/staff\_s.php?eng=1\&id=125}

\author{K. V. Chizh}

\author{V. A. Chapnin}

\author{S. A. Mironov}

\author{V. P. Dubkov}

\author{O. V. Uvarov}

\author{V. P. Kalinushkin}

\affiliation{A.\,M.\,Prokhorov General Physics Institute of the Russian Academy of Sciences, 38 Vavilov Street, Moscow, 119991, Russia}


\author{V.~M.~Senkov}

\affiliation{P.\,N.\,Lebedev Physical Institute of the Russian Academy of Sciences, 53 Leninskiy Avenue, Moscow, 119991, Russia}

\author{O.~Y.~Nalivaiko}


\affiliation{JSC ``Integral''--``Integral'' Holding Management Company,  121A, Kazintsa I.\,P. Street, Minsk, 220108, Belarus}

\author{A. G. Novikau}

\author{P. I. Gaiduk}

\affiliation{Belarusian State University, 4 Nezavisimosti Avenue, 220030, Minsk, Belarus}

\date{\today}%

\begin{abstract}
Platinum silicide Schottky diodes formed  on films of polycrystalline Si doped by phosphorus   are demonstrated to be efficient and manufacturable CMOS-compatible temperature sensors for microbolometer detectors of radiation. 
Thin-film platinum silicide/poly-Si diodes have been produced by a CMOS-compatible process on artificial Si$_3$N$_4$/SiO$_2$/Si(001) substrates simulating the bolometer cells.
Layer structure and phase composition of the original Pt/poly-Si films and the Pt silicide/poly-Si films synthesized by a low-temperature process have been studied by means of the scanning transmission electron microscopy; they have also been explored by means of the two-wavelength X-ray structural phase analysis and the X-ray photoelectron spectroscopy. 
Temperature coefficient of voltage for the forward current of a single diode is shown to reach the value of about $-$2\,\%/{\textcelsius} in the temperature interval from 25 to 50\,{\textcelsius}.
 \end{abstract}



\maketitle

\section{Introduction }

Over the past two decades, there was a qualitative breakthrough in the development of thermal imaging devices based on uncooled IR focal plane arrays (FPAs). At present, two types of bolometer FPAs are commonly used as sensors in uncooled IR imagers: the ones based on vanadium oxide (VO$_x$) or amorphous silicon ($\alpha$-Si:H) thermistors.\cite{Akin_CMOS_Thermal} Despite that the VO$_x$ is not compatible with the CMOS process, the imagers based on this materials are most common on the market. It should be noted that CMOS compatibility is an important advantage of the temperature sensor material: a CMOS-compatible manufacturing process considerably decreases the FPA production cost and simplifies its mass production. The $\alpha$-Si:H FPAs are CMOS compatible. Another type of CMOS-compatible IR sensors are based on poly-SiGe thermistors; they demonstrate high sensitivity and pixel uniformity.\cite{SiGe-Bolometers} Yet they are not used for commercial production of FPAs at present. 

Nowadays, reduction of the FPAs production cost in conjunction with possibility of their mass production is an important technological goal;  solution of this task will lower prices of IR imagers on the market and widen an area of their applications. One of promising approaches to solution of this task consists in development of low-cost diode FPAs manufactured by a process compatible with the CMOS one; there are many potential advantages of this type of bolometer sensors in comparison with the thermistor ones such as less pixel size at the same sensitivity, 
the possibility of increasing the sensitivity due to connection of several diodes in one pixel,
high uniformity of pixels properties,
high stability and low noise,
significantly lower power consumption,
considerably less heating of a sensing element, 
and consequently, the possibility of operation in the permanent current mode or variation of the current pulses duration over a wide range,
flexibility in the choice of operating modes: constant current or constant voltage, forward or reverse biasing, etc., and finally
easier integration into a standard CMOS process.

Some time ago, a novel class of uncooled microbolometer IR FPAs was  developed in which Si-on-insulator (SOI) diodes had been utilized as temperature sensors.{\cite{Kimata_3-Si_FPAs}}
Presently, a format of these FPAs has reached 2 megapixels;\cite{Diode_bolometers-Mitsubishi-camera-2Mpix} their noise equivalent temperature difference (NETD) has made  60 mK at  the frame rate of 15 Hz and the f-number of 1 that is a very good sensitivity for large-format uncooled FPAs.  
A SOI-diode VGA FPA with NETD of 21 mK (f/1, 30 Hz) and an uncooled infrared micro-camera with the same parameters have also been recently demonstrated.\cite{Diode_bolometers-Mitsubishi-camera-2Mpix,Diode_bolometers-Mitsubishi-22mK-camera} 
These remarkable achievements have given a pulse to search for simple CMOS-compatible technological solutions based on diode bolometers which would be suited to mass production of low cost IR FPAs with NETD figures appropriate to various civil---medical and industrial---and tactical applications.\cite{Diode_bolometers-turkish1,*Diode_bolometers-turkish2, Diode_bolometers-1, patent-Schottky-bolometer,*patent-diode-bolometer} A possible attractive solution consists in utilization of  metal/poly-Si Schottky junctions for formation of sets of connected temperature sensors on bolometer membranes;\cite{patent-Schottky-bolometer} the Schottky-barrier bolometers were likely first  proposed theoretically for high-sensitive cooled detectors in Ref.\,\onlinecite{Schottky-Diode_bolometers-cooled}. 

In our recent article, nickel silicide Schottky diodes formed on phosphorus doped polycrystalline Si films were demonstrated to be a promising alternative to  SOI-diodes in monolithic uncooled microbolometer FPAs.\cite{NiSi-bolometer} 
Absolute values of their temperature coefficients of voltage and current were found to reach 0.6\,\%/{\textcelsius} for the forward bias and be around 2.5\,\%/{\textcelsius} for the reverse bias of the diodes despite that the studied diodes were far from perfectness. A relative ease of production as well as  possibility of cascade connection of the Schottky diodes increasing the temperature sensitivity of bolometer elements and application of layers of the diode structures of  bolometer cells as additional absorbers of  the incident radiation  may be obvious practical advantages of utilization of the Ni-silicide/poly-Si junctions.

A solution even more suitable for the industry is considered in the current article. We propose to apply platinum silicide/poly-Si Schottky diodes for temperature sensing in microbolometers. In addition to the mentioned advantages of the nickel silicide/poly-Si diodes,
ease of integration of technological process of the Pt silicide/poly-Si diode formation  into the standard CMOS technology of VLSI manufacturing,\cite{Nano-2014_silicides} 
in analogy with the technology of the monolithic PtSi/Si IR FPAs well developed in the industry,\cite{Murarka,Rogalski_1,*Rogalski_2,*Rogalski_3, *Rogalski_4, PtSi-IrSi_Micelec-our-EN,*PtSi-Lyapunov-1,*PtSi-Lyapunov-2, Kimata_2-PtSi, Kimata_3-Si_FPAs} is the main merit of this approach.

To test the proposed solution we have produced by a CMOS-compatible process the Pt-silicide/poly-Si:P Schottky-diode structures simulating the topology of the prospective sensitive cells of bolometers  and examined them using the scanning transmission electron microscopy, the two-wavelength X-ray structural phase analysis (diffractometry and reflectometry)  and the X-ray photoelectron spectroscopy.
We have also studied the \textit{I--V} characteristics of the structures and investigated their temperature sensitivity (temperature coefficients of voltage $(TCU = U^{-1}dU/dT)$ for different values of the stabilized current) at different temperatures.
We have concluded that Pt silicide/poly-Si Schottky diodes may be considered as a promising solution for utilization as temperature sensors of prospective diode bolometers.

Let us proceed now to the presentation of the obtained results.


\section{Samples and Methods}
\subsection{Production of the Schottky-Diode Structures}

\begin{figure*}[th]
\includegraphics[scale=1]{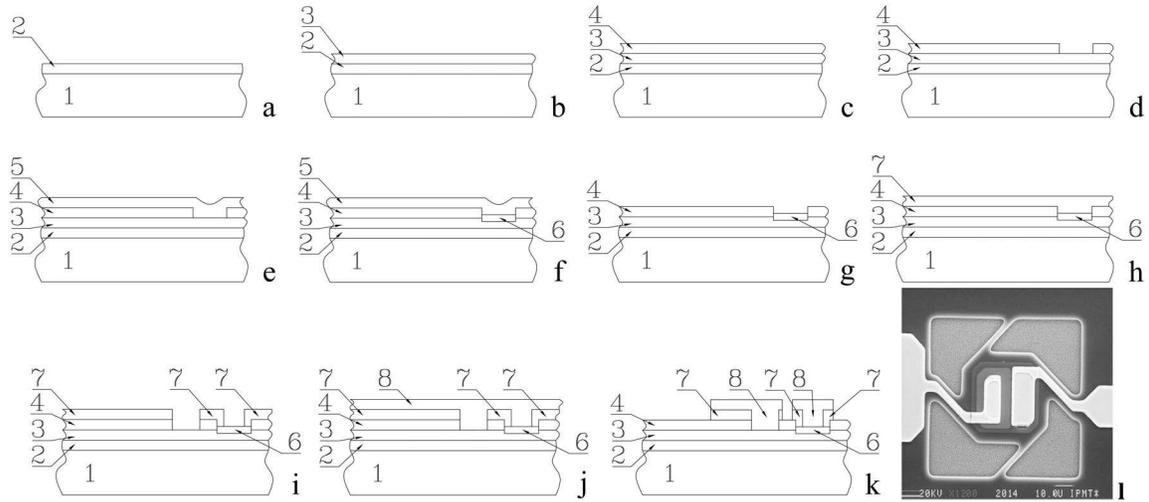}
\caption{\label{fig:process}
A schematic of the process of the bolometer Pt silicide/poly-Si diode formation;
the steps are as follows: 
(a) formation of the bolometer membrane layers:
(1) a CZ Si:B (100) wafer, $\rho=12$\,$\Omega$cm; 
(2) bolometer membrane layers [thermal SiO$_2$ (527\,nm) and pyrolytic Si$_3$N$_4$ (174\,nm)]; 
(b) deposition and doping of the polycrystalline silicon layer:
(3) poly-Si:P (125\,nm, SiH$_4$ thermal decomposition at 620\,{\textcelsius}; P$^+$ implantation at $E_{\rm{P^{^+}}}$=\,60\,keV to the dose of $1.25\times 10^{14}$\,cm$^{-2}$; dopant activation by annealing at 850\,{\textcelsius} for 30\,min);
formation of the diode lateral dimensions:
the contact lithography and plasma etching of poly-Si:P;
(c) deposition of the SiO$_2$ insulator:
(4) SiO$_2$ (400\,nm, PECVD);
(d) formation of a window for the Pt silicide/poly-Si:P junction: 
the contact lithography and plasma etching of SiO$_2$;
(e) deposition of platinum:
(5) Pt film (35\,nm, magnetron sputtering);
(f) formation of the Pt silicide/poly-Si:P junction:
(6) Pt silicide ($\sim$\,60\,nm; formed by annealing at 550\,{\textcelsius} in the mixture of N$_2$ and H$_2$ for 30\,min; $\sim$\,30\,nm thick poly-Si:P layer remains under the Pt silicide film); 
(g) etching of the excess platinum:
warm H$_2$O\,:\,HCl\,:\,HNO$_3$ [4\,:\,3\,:\,1] for 4 min;
(h) deposition of the SiO$_2$ insulator:
(7) SiO$_2$ (300\,nm, PECVD); 
(i) formation of windows for the ohmic contacts to Pt silicide and poly-Si:P:
the contact lithography and plasma etching of SiO$_2$;
(j) deposition of a metal film (W--Ti--Al) for the ohmic contacts to Pt silicide and poly-Si:P:
(8) W--Ti (125\,nm) and Al (450\,nm) film, magnetron sputtering;
(k) formation of contact pads and interconnects:
the contact lithography and plasma etching of the metal film.
(l) A SEM image of the diode: the shape corresponds to a bolometer cell; the left contact is to Pt silicide, the right one is to polysilicon.
}
\end{figure*}

Chips with the Pt silicide/poly-Si Schottky-diode structures  were formed on commercial Czochralski-grown single-crystalline silicon wafers ($\rho = 12\,\Omega$cm, (100)-oriented, $p$-type) coated by a 527\,nm thick layer of SiO$_2$ formed by thermal oxidation and a  174\,nm thick layer of pyrolytic Si$_3$N$_4$ (the dielectric layers simulated a supporting membrane of a bolometer cell).\cite{Orion-2012,NiSi-bolometer} The process details are presented in Fig.\,\ref{fig:process}.
A film of polycrystalline Si with the thicknesses of 125\,nm was deposited on the Si$_3$N$_4$ surface by thermal decomposition of monosilane at the substrate temperature of 620\,{\textcelsius}; then it was  doped by  implantation of phosphorus ions ($E_{\rm{P^{^+}}}$=\,60\,keV) to the dose of $1.25\times 10^{14}$\,cm$^{-2}$  and annealing at 850\,{\textcelsius} for 30 min. Afterward, a diode lateral dimensions were formed by the contact lithography and plasma etching of poly-Si:P. At the next step a silicon dioxide dielectric film, 400\,nm in thickness, was formed by plasma enhanced chemical vapour deposition (PECVD). Then  a window in SiO$_2$ for the Pt silicide/poly-Si:P junction was formed by means of the contact lithography and plasma etching and 
a 35-nm 
thick layer of platinum was deposited by magnetron sputtering at room temperature. 
Platinum silicide was formed by annealing at 550\,{\textcelsius} for 30\,min in the gas mixture of N$_2$ and H$_2$; the Pt silicide/poly-Si:P junction, a $\sim$\,60-nm Pt-silicide  layer overlying $\sim$\,30-nm poly-Si:P layer (Pt polycide), formed after this step. 
Excess platinum was removed by chemical etching in warm H$_2$O\,:\,HCl\,:\,HNO$_3$ [4\,:\,3\,:\,1] for 4 min. 
An additional SiO$_2$ insulator with the thickness of 300\,nm was deposited by PECVD. Windows for ohmic contacts to Pt silicide and poly-Si:P were formed by the contact lithography and plasma etching of SiO$_2$. At the final steps a metal film of W--Ti (125\,nm) and Al (450\,nm)  was deposited by magnetron sputtering and ohmic contacts  to Pt silicide and poly-Si:P were formed; interconnects and contact pads  were formed by the contact lithography and plasma etching of the metal film.

Finally, chips were cut from the wafers (Fig.\,\ref{fig:process}\,l) and placed into housings which enabled the study of electrical properties of the samples at any temperatures from cryogenic to over 100\,\textcelsius.
Golden wires were welded to the contact pads for electrical measurements.

Satellites were produced at the steps of crucial treatments for studies of structure and composition of the formed layers and examined by means of the transmission electron microscopy; some satellites (Fig.\,\ref{fig:process}\,e,\,f) were investigated by the two-wavelength X-ray structural phase analysis and the X-ray photoelectron spectroscopy.

\subsection{Experimental Techniques and Instruments} 

Structural perfectness and chemical composition of the layers were explored by means of the scanning transmission electron microscopy (STEM) using the produced satellites. The Carl Zeiss Libra-200 FE HR transmission electron microscope was used. The WSxM software was applied for image processing.\cite{WSxM}

For the complete and reliable determination of the Schottky junction structure the  two-wavelength X-ray diffractometry and reflectometry methods were applied.\cite{Touryanski-patent,*Touryanski-article} 
The method of the two-wavelength X-ray structural phase analysis comprises  measurements of the scattering or reflectance angular diagrams at two wavelengths simultaneously for a single scan that essentially increases the determination accuracy of parameters  of  studied objects. It enables the analysis of multilayer structures with diffused  interfaces and permits the quantitative analysis of X-ray reflectometry data down to zero grazing  angle.\cite{Nanostructures-2013_X-rays}  

The  analysis was performed using the CompleXRay-C6 instrument at Cu\,K$_{\alpha}$ ($\lambda = 0.154$\,nm) and Cu\,K$_{\beta}$ ($\lambda = 0.139$\,nm) bands;\cite{Touryanski-patent,*Touryanski-article} the scanning steps of the diffraction patterns $\Delta(2\theta) = 0.10^{\circ}$. 
PDF-2 Powder Diffraction Database (Joint Committee on Powder Diffraction Standards\,--\,International Centre for Diffraction Data) was used for the phase composition analysis.

X-ray reflectance angular diagrams were scanned at the total external reflection; the $\theta$--2$\theta$ geometry was used.
In the used instrument, a determination of the structure parameters (fitting of the experimental curves) is carried out semi-automatically by using the genetic algorithm.\cite{X-rays_Wormington}
The reflectogram calculation is performed using  the recurrence relations\cite{X-ray_recurrent} in which 
the N{\'{e}}vot--Croce metod\cite{X-rays_Nevot&Croce, X-rays_Wormington} is introduced for taking account of the layer roughness contribution (or widths of interfaces). If the interface layers are of diffusion nature their  widths may reach several nanometers; it is hard to speak about roughness in this case and we use a term ``the  N{\'{e}}vot--Croce parameter''.

The satellites obtained at two crucially important steps of the diode production (Fig.\,\ref{fig:process}\,e,\,f) were explored by means of the X-ray structural analysis. 

The surface layers of the satellites of the structures before and after the silicide formation (Fig.\,\ref{fig:process}\,e,\,f) were studied by means of the  X-ray photoelectron spectroscopy (XPS).
The measurements were carried out using a cylindrical mirror electron energy analyser\cite{Cylindrical_Mirror_Analyzer} (Riber\,EA\,150) installed in the ultrahigh-vacuum analytical chamber of the Riber SSC2 surface science center; the residual gas pressure in the analytical chamber did not exceed 10$^{-7}$\,Pa  during the experiments.  
Non-monochromatic
Al\,K$_{\alpha}$ X-rays ($h\nu = 1486.7$\,eV) were used for photoexcitation of electrons. 
Survey spectra were scanned at the resolution (FWHM) better than 1.8\,eV; 
high resolution spectra of specific elements were obtained at the resolution not worse than 0.96\,eV.
XPSPEAK\,4.1 XPS peak fitting program was utilized for treatment of the photoelectron spectra.
The same degree of asymmetry was used in the deconvolution of the platinum peaks. Relative concentrations of atoms were estimated from ratios of normalized areas under corresponding peaks.  Shifts of peaks related to elements in chemical compounds were compared with the NIST X-ray Photoelectron Spectroscopy (XPS) Database.\cite{XPS_NIST_Database}
The samples were washed in the ammonia-peroxide solution (NH$_4$OH\,:\,H$_2$O$_2$\,:\,H$_2$O $[1:1:4]$, boiling for 5 min), rinsed by the deionized water and the warm isopropanol, and dried in the clean air before the analysis.

{\it I--V} characteristics of the Schottky diodes were measured in darkness at different temperatures varied in the range from 25 to 50\textcelsius. 
The values of the temperature coefficient of voltage (\textit{TCU}) were derived from linear fits of the temperature dependences of logarithms of voltage measured at different values of the stabilized current through the diodes: $TCU = d[\ln U(T)]/dT$ ($TCU= d[\ln R(T)]/dT = TCR$ for the current $I$ independent of the temperature $T$; as usually, $R$ is the resistance and \textit{TCR} is the temperature coefficient of resistance). The temperature was stabilized in the thermostat with an accuracy of about 0.1\,\textcelsius.

\begin{figure*}[th]
\includegraphics[scale=1]{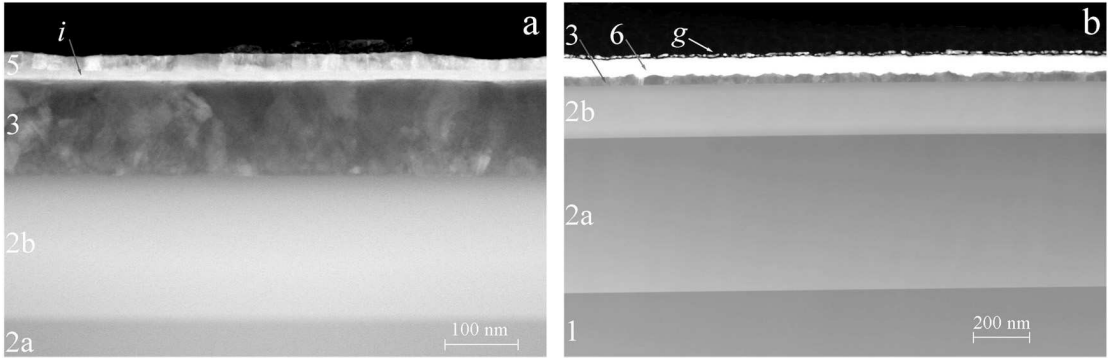}
\caption{\label{fig:STEM}
STEM images of the Pt/Si  [Fig.\,\ref{fig:process}\,e] (a) and Pt silicide/Si [Fig.\,\ref{fig:process}\,f] (b) samples; layer numbering corresponds to that in Fig.\,\ref{fig:process}: 
(1) a CZ Si:B wafer;
(2) bolometer membrane layers:
(2a) SiO$_2$ and 
(2b) Si$_3$N$_4$;
(3) poly-Si:P;
(5) Pt film, a dense interface layer ($i$) is observed beneath the layer of polycrystalline Pt;
(6) Pt silicide, a surface layer ($g$) composed of grains of a dense substance  is seen to cover this layer.
}
\end{figure*}

\begin{figure}[th]
\includegraphics[scale=.95]{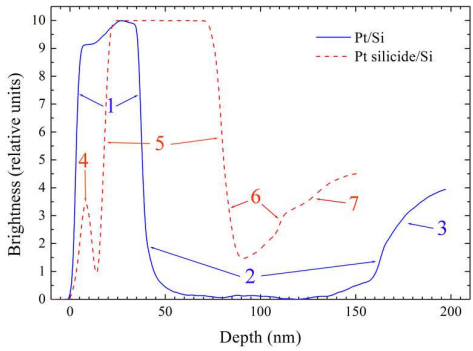}
\caption{\label{fig:STEM-profiles}(Color online)
Brightness profiles of the STEM images of the Pt/Si and Pt silicide/Si samples (Fig.\,\ref{fig:STEM}) obtained by averaging the images over axes running along the layers: 
(1) Pt, $\sim$\,32\,nm, with an interface layer;
(2) poly-Si, $\sim$\,125\,nm;
(3) Si$_3$N$_4$;
(4) a granular surface layer of unknown composition;
(5) Pt silicide, $\sim$\,66\,nm;
(6) poly-Si, $\sim$\,32\,nm;
(7) Si$_3$N$_4$.
}
\end{figure}

\section{Characterization and silicide formation analysis}

\subsection{Transmission Electron Microscopy}


STEM micrographs obtained from the structures before and after annealing for the platinum silicide layer formation (Fig.\,\ref{fig:process}\,e,\,f) are presented in Fig.\,\ref{fig:STEM}. They demonstrate that the Pt/poly-Si structure (Fig.\,\ref{fig:STEM}\,a) contains, in addition to polycrystalline silicon and platinum layer, an  interfacial layer of unresolved internal structure between the Si and Pt layers. (The EDX microanalysis indicates the presence of Si in the interfacial layer.) In the Pt silicide/poly-Si structure (Fig.\,\ref{fig:STEM}\,b), besides the polysilicon and silicide layers, a granular  layer  consisting of a relatively dense substance lays atop the silicide.

The images in Fig.\,\ref{fig:STEM} allowed us to roughly estimate the composition of the platinum silicide layer using the mass conservation law just like it had been done in Ref.\,\onlinecite{NiSi-bolometer} for the nickel silicide Schottky diodes. We measured the mean thicknesses of the Pt and poly-Si layers before and Pt-silicide and poly-Si layers after the Schottky junction formation (Fig.\,\ref{fig:STEM-profiles}).  
In the structure before formation of the Schottky junction, the average thicknesses of the layer 1 and the layer 2 in Fig.\,\ref{fig:STEM-profiles}  were evaluated as  nearly 32 and about 125\,nm, respectively. The layer 2 corresponds to poly-Si whereas the layer 1 consists of two clearly resolved sublayers, which we could attribute to Pt and some compound of Pt and Si, with the approximate thicknesses of 21 to 27\,nm for Pt and, respectively, 11 to 5\,nm for the Pt--Si compound.
The average thicknesses of the Pt-silicide and poly-Si layers after formation of the Schottky barrier were estimated as approximately 66 and 32\,nm, respectively.
 
By fitting the data (the density of poly-Si is adopted to be 2.0\,g/cm$^3$, that of the initial Pt film is assumed to be equal to 21.45\,g/cm$^3$ at room temperature)\cite{Physical_Quantities-Handbook} we obtain the best agreement with the observation if Pt--Si\,=\,Pt$_2$Si (the density is 16.268\,g/cm$^3$)\cite{Silicides} and consequently the density of the resultant silicide is approximately equal to 12.83\,g/cm$^3$ for the initial Pt silicide sublayer thickness of 5\,nm and 12.35\,g/cm$^3$ for the as-deposited Pt silicide sublayer thickness of 11\,nm (when fitting we assumed that the initial Pt silicide sublayer consists of some known Pt silicide phase, see, e.\,g., Refs.~\onlinecite{Ni3Si2_&_Pt6Si5,*PtSi_Pt2Si_Pt6Si5_Pt3Si_Pt7Si3,Silicides,Murarka}; the density of PtSi is adopted to be 12.394\,g/cm$^3$).

As a result we can make a preliminary conclusion about the composition of the resultant Pt silicide layer: it can be composed  either entirely by PtSi or, more likely, by a combination of PtSi and Pt$_2$Si. The maximum content of Pt$_2$Si  is estimated as approximately 11\,vol.\% (14\,wt.\%) for the silicide density of 12.83\,g/cm$^3$. Unfortunately, we cannot resolve from our STEM measurements whether this layer consists of a mixture of the PtSi and Pt$_2$Si grains or it is composed by the PtSi and Pt$_2$Si layers. We belive however that it likely should be layered since the initial structure is layered and the components diffuse between relatively homogeneous layers during the solid-phase reaction of the  silicide formation. 

And finally, we should mention that an average thickness of the uppermost granular layer of the Pt silicide/poly-Si structure (Fig.\,\ref{fig:STEM-profiles}, number 4) is about 15\,nm; its composition cannot be determined from the STEM images.

\begin{figure}[t]
\includegraphics[scale=.95]{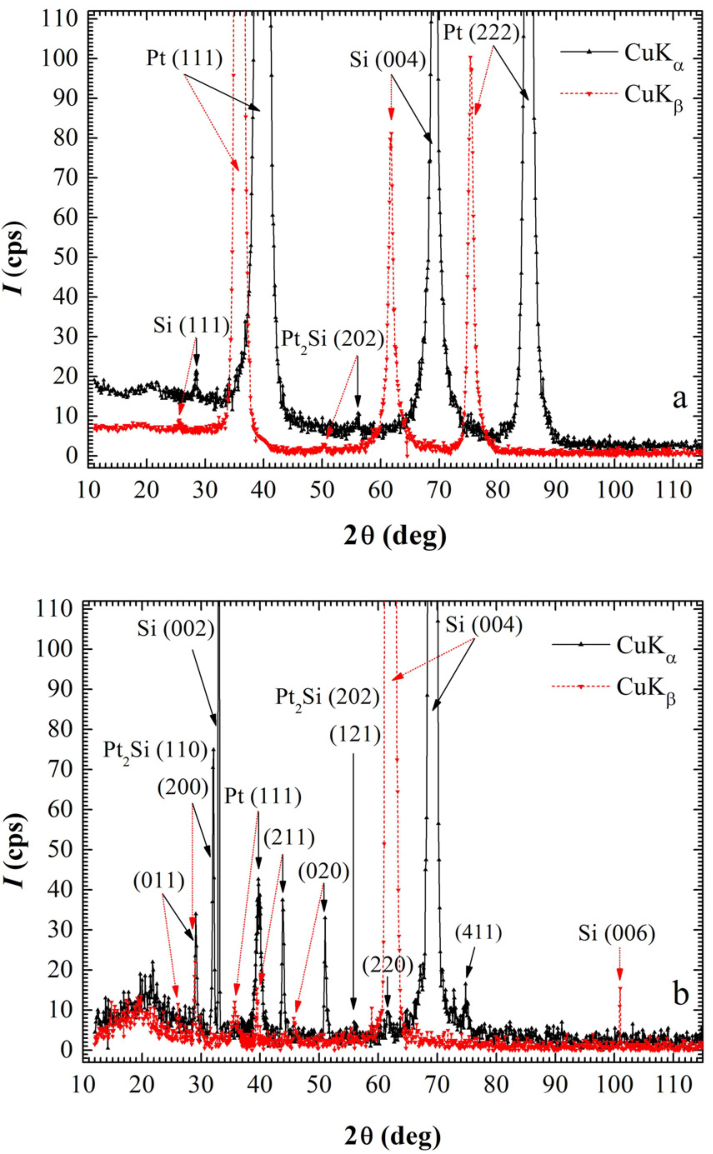}
\caption{\label{fig:diffract}(Color online)
X-ray diffraction patterns of the  Pt/poly-Si (a) and Pt~silicide/poly-Si (b) multilayer structures measured for Cu\,K$_{\alpha}~(\lambda =0.154$\,nm) and Cu\,K$_{\beta}~(\lambda =0.139$\,nm) lines; 
the curves in the panel (a) were measured in the $(\theta - 0.3^{\circ})$--$2\theta$ mode, 
the curves in the panel (b) were measured in the $\theta$--$2\theta$ mode; Miller indices without substance chemical formulas  relate to PtSi;  Pt$_2$Si(110) and PtSi(200), Pt$_2$Si(110) and PtSi(121) reflexes pairwise overlap in the panel (b);
a list of reflexes is given in Table\,\ref{tab:diffract}.
}
\end{figure}

\subsection{X-ray Structural Analysis}

\subsubsection{X-ray Diffractometry}

\begin{table*}
\caption{\label{tab:diffract}
Reflexes  observed in the X-ray diffraction patterns  from the  Pt/poly-Si structure (Fig.\,\ref{fig:diffract}\,a) and Pt~silicide/poly-Si structure (Fig.\,\ref{fig:diffract}\,b). 
}
\begin{ruledtabular}
\begin{tabular}{lcccclccl}
\multicolumn{2}{c}{Phase}& Diffraction &   \multicolumn{3}{c}{Cu\,K$_\alpha$}&\multicolumn{3}{c}{Cu\,K$_\beta$}\\
Composition& Crystal System& Crystal Plane &\multicolumn{2}{c}{$2\theta$}
&  Band &\multicolumn{2}{c}{$2\theta$} 
& Band \\
& &    Miller Indices  & Tabulated\footnote{PDF-2 Powder Diffraction Database, Joint Committee on Powder Diffraction Standards\,--\,International Centre for Diffraction Data.} & Measured & Intensity & Tabulated$^{\rm a}$ & Measured&  Intensity\\ 
&&$(h~k~l)$&  \multicolumn{2}{c}{(deg)} 
& (cps) &\multicolumn{2}{c}{(deg)}
&  (cps)\\
\hline\\
%
%
\multicolumn{9}{c}{Pt/poly-Si (Fig.\,\ref{fig:diffract}\,a)}\\ \\
Si& cubic &(1\,1\,1) &\multicolumn{1}{d}{28.44} & \multicolumn{1}{d}{28.43} & \multicolumn{1}{d}{7} & \multicolumn{1}{d}{25.65} & \multicolumn{1}{d}{25.61} & \multicolumn{1}{d}{2}\\
Pt\footnote{The peaks of Pt in the Pt/poly-Si structure are shifted with respect to their tabulated positions; this shift may be a result of admixing from 3 to 5\,at.\,\% of Si to Pt.}
& cubic &(1\,1\,1)& \multicolumn{1}{d}{39.79} & \multicolumn{1}{d}{39.55} & \multicolumn{1}{d}{8038} & \multicolumn{1}{d}{35.73} & \multicolumn{1}{d}{35.6} & \multicolumn{1}{d}{2241}\\
Pt$_2$Si& tetragonal&(2\,0\,2) & \multicolumn{1}{d}{56.23} & \multicolumn{1}{d}{56.2} & \multicolumn{1}{d}{5} & \multicolumn{1}{d}{50.30} & \multicolumn{1}{d}{50.2} & \multicolumn{1}{d}{2}\\
Si& cubic&(0\,0\,4)  & \multicolumn{1}{d}{69.13} & \multicolumn{1}{d}{69.2} & \multicolumn{1}{d}{228} & \multicolumn{1}{d}{61.69} & \multicolumn{1}{d}{61.7} & \multicolumn{1}{d}{81}\\
Pt$^{\rm b}$
& cubic&(2\,2\,2) & \multicolumn{1}{d}{85.80} & \multicolumn{1}{d}{85.3} & \multicolumn{1}{d}{368} & \multicolumn{1}{d}{75.71} & \multicolumn{1}{d}{75.4} &\multicolumn{1}{d}{100}\\ \\
%
%
\multicolumn{9}{c}{Pt~silicide/poly-Si (Fig.\,\ref{fig:diffract}\,b)}\\ \\
PtSi& orthorhombic&(0\,1\,1) & \multicolumn{1}{d}{29.10} & \multicolumn{1}{d}{29.1} &\multicolumn{1}{d}{34} & \multicolumn{1}{d}{26.18}  & \multicolumn{1}{d}{26.1}  & \multicolumn{1}{d}{12} \\
PtSi\,\& & orthorhombic&(2\,0\,0)\\
Pt$_2$Si& tetragonal&(1\,1\,0) & \multicolumn{1}{d}{32.15} & \multicolumn{1}{d}{32.1} & \multicolumn{1}{d}{75} & \multicolumn{1}{d}{28.91}  & \multicolumn{1}{d}{29.0}  &  \multicolumn{1}{d}{22}\\
Si& cubic&(0\,0\,2)& \multicolumn{1}{d}{32.96} & \multicolumn{1}{d}{33.0} & \multicolumn{1}{d}{298} & \multicolumn{1}{d}{29.97}  & \multicolumn{1}{d}{29.8}  & \multicolumn{1}{d}{5} \\
Pt& cubic&(1\,1\,1) & \multicolumn{1}{d}{39.76} & \multicolumn{1}{d}{39.73} & \multicolumn{1}{d}{38} & \multicolumn{1}{d}{35.73}  & \multicolumn{1}{d}{35.7}  & \multicolumn{1}{d}{12} \\
PtSi& orthorhombic&(2\,1\,1) & \multicolumn{1}{d}{43.91} & \multicolumn{1}{d}{43.9} & \multicolumn{1}{d}{34} &  \multicolumn{1}{d}{39.39} & \multicolumn{1}{d}{39.45}  & \multicolumn{1}{d}{15}\\
PtSi& orthorhombic&(0\,2\,0) & \multicolumn{1}{d}{50.91} & \multicolumn{1}{d}{51.1} & \multicolumn{1}{d}{33} & \multicolumn{1}{d}{45.60}  & \multicolumn{1}{d}{45.7}  & \multicolumn{1}{d}{8} \\
PtSi& orthorhombic&(1\,2\,1) & \multicolumn{1}{d}{56.06} & \multicolumn{1}{d}{56.1} & \multicolumn{1}{d}{5} & \multicolumn{1}{d}{50.13}  & \multicolumn{1}{d}{50.1}  & \multicolumn{1}{d}{1} \\
Pt$_2$Si& tetragonal&(2\,0\,2) & \multicolumn{1}{d}{56.23} & \multicolumn{1}{d}{56.2} & \multicolumn{1}{d}{3} & \multicolumn{1}{d}{50.30}  & \multicolumn{1}{d}{50.4}  & \multicolumn{1}{d}{2} \\
PtSi& orthorhombic &(2\,2\,0)& \multicolumn{1}{d}{61.50} & \multicolumn{1}{d}{61.5} & \multicolumn{1}{d}{5} & \multicolumn{1}{d}{54.90}  & \multicolumn{1}{d}{55.1}  & \multicolumn{1}{d}{2}\\
Si& cubic&(0\,0\,4) & \multicolumn{1}{d}{69.13} & \multicolumn{1}{d}{69.1} & \multicolumn{1}{d}{75390} & \multicolumn{1}{d}{61.69}  & \multicolumn{1}{d}{61.7}  & \multicolumn{1}{d}{40870} \\
PtSi& orthorhombic&(1\,1\,4) & \multicolumn{1}{d}{70.85} & \multicolumn{1}{d}{79.9} & \multicolumn{1}{d}{7} & \multicolumn{1}{d}{63.01}  &  \multicolumn{1}{d}{ } & \multicolumn{1}{d}{ } \\
Unidentified & &&  & \multicolumn{1}{d}{72.8} &  \multicolumn{1}{d}{15} &  &  &  \\
PtSi& orthorhombic&(4\,1\,1) & \multicolumn{1}{d}{74.91} & \multicolumn{1}{d}{74.8} & \multicolumn{1}{d}{17} & \multicolumn{1}{d}{66.50}  & \multicolumn{1}{d}{66.5}  & \multicolumn{1}{d}{4} \\
PtSi& orthorhombic&(0\,3\,1) & \multicolumn{1}{d}{82.25} &  \multicolumn{1}{d}{82.3} & \multicolumn{1}{d}{7} & \multicolumn{1}{d}{72.73}  &   &  \\
Si& cubic&(0\,0\,6) & \multicolumn{1}{d}{116.80}  &   &   & \multicolumn{1}{d}{100.54}  &\multicolumn{1}{d}{100.7}  & \multicolumn{1}{d}{16} \\
\end{tabular}
\end{ruledtabular}
\end{table*}

For determination of composition of the Pt/poly-Si and Pt silicide/poly-Si structures their qualitative X-ray structural phase analysis was performed. Diffraction patterns of the studied samples for Cu\,K$_{\alpha}$ and Cu\,K$_{\beta}$ lines are shown in Fig.\,\ref{fig:diffract}. To reduce the signal from the single-crystalline Si substrate when studying the Pt/poly-Si sample the $(\theta - 0.3^{\circ})$--$2\theta$ curves were measured instead of the $\theta$--$2\theta$ ones (the $\theta$ angle was decreased by $0.3^{\circ}$ during scanning). Comparison of the  curves obtained in these modes shows that the intensity and positions of reflexes from polycrystalline phases remains unchanged for the both  scanning modes.

In the Pt/poly-Si structure (Fig.\,\ref{fig:diffract}\,a), intense reflexes from platinum---Pt(111) and Pt(222)---and the Si(111) reflexes from the polycrystalline  silicon are observed.  The attenuated, although relatively strong,  Si(004) reflexes from the single-crystalline silicon substrate are also detected.  In addition, relatively weak peaks corresponding to the Pt$_2$Si phase are revealed; the latter phase turned out to appear during the   deposition of Pt by magnetron sputtering at room temperature.  

Intense peaks related to the PtSi phase are observed in the Pt silicide/poly-Si structure (Fig.\,\ref{fig:diffract}\,b). Additionally, Pt(111) lines are also registered which indicates that some amount of unreacted platinum still remains in the structure. The presence of reflexes from Pt$_2$Si also is  not excluded but they seem to overlap  with the  intense lines of PtSi. The intense Si(002) and Si(004) reflexes are present in the patterns; the Si(006) reflex is seen in the range of high diffraction angles in the curve for Cu\,K$_{\beta}$.
It should be noticed that the Si(111) peak of the polycrystalline silicon at $2\theta = 28.44^{\circ}$, which is observed in the Pt/poly-Si structure, is absent in the diffraction pattern of the Pt silicide/poly-Si structure although the polysilicon layer is well seen in the STEM image of this sample (Fig.\,\ref{fig:STEM}\,b). 

The observed reflexes for the both structures and corresponding phases of substances are listed in Table\,\ref{tab:diffract}.


\subsubsection{X-ray Reflectometry}

\begin{figure}[th]
\includegraphics[scale=.95]{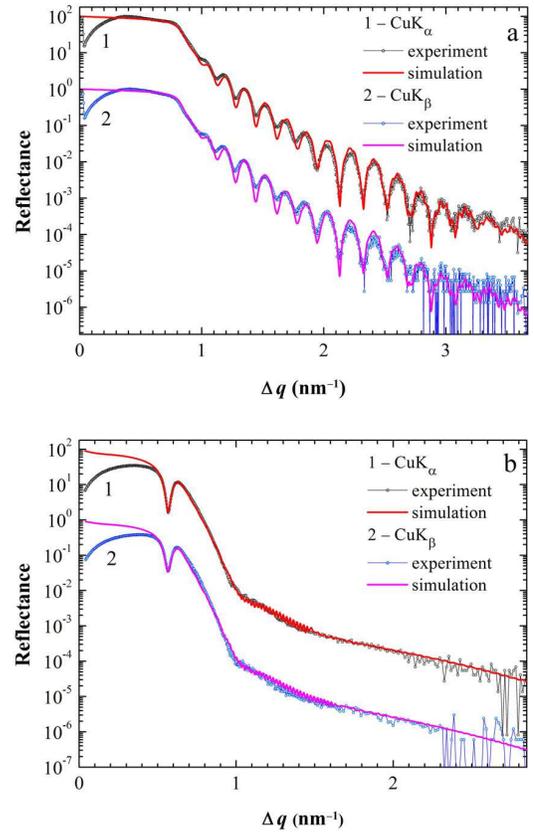}
\caption{\label{fig:reflect}(Color online)
X-ray specular reflectance of the  Pt/poly-Si (a) and Pt silicide/poly-Si (b) multilayer structures;  $\Delta q = 4\pi\sin\theta/\lambda$ is the change in the wave number of the reflected quanta in the incidence plane; experimental data for the Cu\,K$_\alpha$ and Cu\,K$_\beta$ lines are presented by dots connected by thin lines, the thicker solid lines represent the simulated curves; the curves 1 are multiplied by 100.
}
\end{figure}

\begin{table}
\caption{\label{tab:reflect}A sequence of layers in the Pt/poly-Si structure before platinum silicide formation and the Pt silicide/poly-Si structure after platinum silicide formation and their parameters; the data are derived from the X-ray reflectogram.}
\begin{ruledtabular}
\begin{tabular}{llcccc}
Layer  & Composition  & Roughness\footnote{ The {N{\'{e}}vot--Croce parameter $\sigma$, Ref.\,\onlinecite{X-rays_Wormington}}.}  & Thickness  & Density \\
No. & atomic (wt.\,\%)  &  (nm) &  (nm) &  (g/cm$^3$)\\
\hline\\
%
%
\multicolumn{5}{c}{Pt/poly-Si (Fig.\,\ref{fig:reflect}\,a)}\\ \\
1 & Pt  & \multicolumn{1}{d}{0.8} & \multicolumn{1}{d}{16.2} & \multicolumn{1}{d}{21.45} \\
1a & Pt$_{95}$Si$_5$  & \multicolumn{1}{d}{0.5} & \multicolumn{1}{d}{10.0} & \multicolumn{1}{d}{20.5} \\
2 & Pt$_2$Si  & \multicolumn{1}{d}{0.7} & \multicolumn{1}{d}{5.6} & \multicolumn{1}{d}{16.27} \\
3 & Si, poly& \multicolumn{1}{d}{0.6} & \multicolumn{1}{d}{125} & \multicolumn{1}{d}{2.0} \\
4 & Si$_3$N$_4$ & \multicolumn{1}{d}{0.2} & \multicolumn{1}{d}{200} & \multicolumn{1}{d}{3.1} \\
5 & SiO$_2$ & \multicolumn{1}{d}{0.3} & \multicolumn{1}{d}{550} & \multicolumn{1}{d}{2.2} \\
& Si, mono& \multicolumn{1}{d}{0.3} & substrate & \multicolumn{1}{d}{2.33} \\\\
%
%
\multicolumn{5}{c}{Pt silicide/poly-Si (Fig.\,\ref{fig:reflect}\,b)}\\ \\
1& Pt$_{37}$Si$_{63}$ (Pt$_{80}$Si$_{20}$) & \multicolumn{1}{d}{6.6} & \multicolumn{1}{d}{7.7} & \multicolumn{1}{d}{9.4} \\
2& Pt$_{29}$Si$_{71}$ (Pt$_{74}$Si$_{26}$) & \multicolumn{1}{d}{6.0} & \multicolumn{1}{d}{8.3} & \multicolumn{1}{d}{7.9} \\
3& Pt$_{60}$Si$_{40}$ (Pt$_{91}$Si$_{9}$) & \multicolumn{1}{d}{2.7} & \multicolumn{1}{d}{3.6} & \multicolumn{1}{d}{13.8} \\
4& Pt$_{64}$Si$_{36}$ (Pt$_{92.5}$Si$_{7.5}$) & \multicolumn{1}{d}{2.2} & \multicolumn{1}{d}{3.9} & \multicolumn{1}{d}{14.6} \\
5& PtSi & \multicolumn{1}{d}{1.3} & \multicolumn{1}{d}{52} & \multicolumn{1}{d}{12.4} \\
6& Si, poly & \multicolumn{1}{d}{3.4} & \multicolumn{1}{d}{32} & \multicolumn{1}{d}{2.0} \\
7& Si$_3$N$_4$ & \multicolumn{1}{d}{2.2} & \multicolumn{1}{d}{190} & \multicolumn{1}{d}{3.1} \\
8& SiO$_2$  & \multicolumn{1}{d}{0.4} & \multicolumn{1}{d}{1.7} & \multicolumn{1}{d}{1.9} \\
9& SiO$_2$  & \multicolumn{1}{d}{0.3} & \multicolumn{1}{d}{550} & \multicolumn{1}{d}{2.2} \\
& Si, mono &\multicolumn{1}{d}{0.3} &substrate & \multicolumn{1}{d}{2.33} \\
\end{tabular}
\end{ruledtabular}
\end{table}

X-ray analysis also allowed us to obtain more detailed data on the structural and chemical composition of the Schottky junction layers from the experiments on X-ray reflection. For determination of the layer-by-layer structure of the samples reflectograms (angular dependences of the reflectance) of the  Pt/poly-Si structure and  the Pt silicide/poly-Si structure have been measured (Fig.\,\ref{fig:reflect}; $\Delta q = 4\pi\sin\theta/\lambda$ is the change in the wave number of the reflected quanta in the incidence plane).
Fitting the experimental curves by numerical simulations we have determined the layer composition, the thickness of the layers and their roughness (the N{\'{e}}vot--Croce parameter).\cite{X-rays_Wormington} The obtained data for the both structures are presented in Table\,\ref{tab:reflect}.

The Pt/poly-Si sample contains layers of Pt and Pt$_2$Si with the total thickness of about 31.8\,nm and densities coinciding with the corresponding tabulated values.\cite{Physical_Quantities-Handbook,Silicides} (The Pt layer is subdivided in two sublayers, a layer of pure Pt, layer 1, and a layer of Pt with admixture of 5 at.\,\% of Si, layer 1a.)  The layer thicknesses of poly-Si, Si$_3$N$_4$ and SiO$_2$ correspond to the process parameters (Fig.\,\ref{fig:process}) and their densities coincide with the tabulated values.\cite{Physical_Quantities-Handbook, Kikoin} Note that Pt, Pt$_2$Si and poly-Si are observed by the X-ray phase analysis. The layers of the silicon nitride and silicon dioxide are not observed in the X-ray phase analysis, probably, because they are in the amorphous state.
All the obtained data are in good agreement with the STEM observations; the total thickness of the Pt--Pt$_2$Si layer (31.8\,nm) matches that determined from the STEM images, the thickness of the Pt$_2$Si layer (5.6\,nm) is close to the minimum estimate obtained above from STEM and the mass conservation law. So, we can conclude that the  Pt$_2$Si layer  is really present in the Pt/poly-Si structure directly after Pt deposition, i.\,e. it likely forms during magnetron sputtering of Pt on poly-Si at room temperature.

Numerical simulation of the Pt silicide/poly-Si structure yields four layers composed by Pt and Si in different proportions located on the surface of this structure; their total thickness is about 23\,nm. Then the PtSi layer lies the density of which (12.4\,g/cm$^3$) corresponds to the tabulated one.\cite{Silicides, Murarka} Poly-Si, silicon nitride, silicon dioxide and Si substrate follow PtSi; their densities coincide with the tabulated ones.\cite{Physical_Quantities-Handbook, Kikoin}  

A composition of the first four layers of this structure (layers 1 to 4 in Table\,\ref{tab:reflect}) was calculated from their mean densities determined from the reflectometry data. Their density was assumed to linearly depend on composition in atomic \%. This assumption is proved by the fact that platinum silicides of known composition follow this dependence with a good accuracy (the deviation is less than 7\,\%): e.\,g., PtSi and Pt$_2$Si have the densities of 12.39 and 16.27\,g/cm$^3$, respectively. 

The first two layers appear to be very rough: their N{\'{e}}vot--Croce parameters are comparable with their widths. These layers consist of platinum silicides, silicon and,  according to the X-ray phase analysis, of platinum (see Table\,\ref{tab:reflect}); i.\,e. they are nonuniform in the composition in the lateral direction. Their total thickness determined from the numerical simulation (15.8\,nm) is very close to the thickness of the granular layer determined from STEM (15\,nm). So, we suppose that these layers form the surface granular layer observed by STEM.

The next two layers, according the numerical simulation, are also rough, their total thickness is 7.5\,nm (Table\,\ref{tab:reflect}); they consist of platinum silicides, silicon and maybe also of platinum. However their composition is rather close to that of Pt$_2$Si, they resemble a Pt$_2$Si granular layer with some additional amount of Si. This allows us to identify them as a replacement of the resultant Pt$_2$Si layer deduced above from STEM and the mass conservation law. The total thickness of these layers and the PtSi layer is 59.5\,nm that is close to the mean total  thickness  of the Pt silicide determined  from STEM. A fraction of this presumably Pt$_2$Si--Si compound in the resultant Pt silicide makes 12.6\,vol.\% (14.2\,wt.\%) that is also rather close to the fraction determined from STEM and the mass conservation law.
\footnote{The fraction of the Pt$_2$Si in the Pt silicide/poly-Si structure determined from STEM and the mass conservation law for the initial Pt$_2$Si sublayer thickness of 5.6\,nm in the Pt/poly-Si structure makes approximately 9.9\,vol.\% (around 12.6\,wt.\%).} Some reduction of the PtSi content may be a result of presence of the first two granular layers which were not taken into the account in the estimation made using the mass conservation law.

Poly-Si detected in this structure using the specular reflection is not seen in the diffraction patterns likely because of its small amount. A thin layer of SiO$_2$ of decreased density (1.9\,g/cm$^3$) is situated  between Si$_4$N$_4$ and SiO$_2$ layers with the standard tabulated densities. The presence of this layer seems to be conditioned by the technological features of the structure preparation.
The layers of the silicon nitride and silicon dioxide are not observed in the X-ray phase analysis of this structure likely because Si$_4$N$_4$ and SiO$_2$  are  amorphous in them.

Notice also that small values of reflectance of the latter structure in comparison with the calculated values at $\Delta q < 0.5$\,nm$^{-1}$ (at low grazing angles) are likely conditioned by the nonuniformity of the structure surface.

\begin{figure}[t]
\includegraphics[scale=.9]{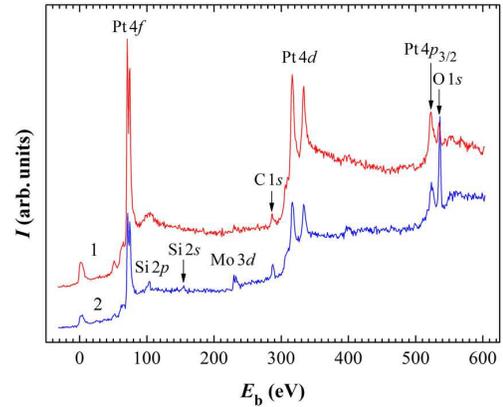}
\caption{\label{fig:XPS}(Color online) 
 Survey XPS spectra of the  Pt/poly-Si (curve 1) and Pt silicide/poly-Si (curve 2)  structures. 
}
\end{figure}

\begin{figure}[t]
\includegraphics[scale=1]{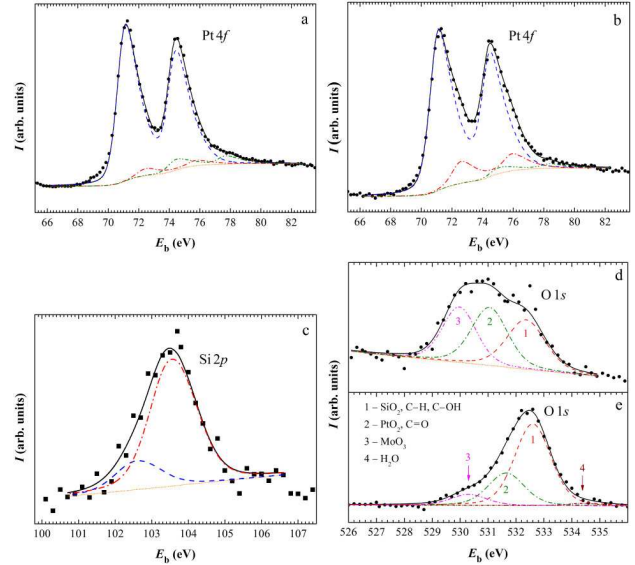}
\caption{\label{fig:XPS-HR}(Color online) 
 High-resolution XPS spectra of Pt (4$f$ band), Si (2$p$ band) and O (1$s$ band) in the top layer of the  Pt/poly-Si (a,\,d) and Pt silicide/poly-Si (b,\,c,\,e)  structures; measured bands are shown by solid lines, peaks obtaind as a result of deconvolution are drawn by dotted and dashed lines; groups of overlapping peaks are marked by common numbers in the panels (d) and (e). 
}
\end{figure}

\subsection{X-ray Photoelectron Spectroscopy}

The XPS analysis of the Pt/poly-Si and Pt silicide/poly-Si structures allowed us to study in more details the chemical composition of the uppermost layers of the both structures and especially the composition of the granular layer in Pt silicide/poly-Si.

In the survey spectra of the Pt/poly-Si sample (Fig.\,\ref{fig:XPS}, curve\,1), photoelectron peaks of Pt dominate; in addition, O\,1$s$ and C\,1$s$ signals are also observed. The Pt\,4$f$ peak (Fig.\,\ref{fig:XPS-HR}\,a) is composed by three components: a peak with the 
lowest electron binding energy ($E_{\mathrm{b}}=71.1$\,eV) corresponds to metallic Pt, the next peak shifted to greater $E_{\mathrm{b}}$ (72.5\,eV) is associated with the Pt$_2$Si compound, and $E_{\mathrm{b}}$ of the last peak (74.4\,eV) is close to that of Pt dioxide. Fractions of Pt atoms in each of the mentioned states make up a  proportion of $90:4:6$, respectively. A fraction of platinum silicide in the film is so small that the Si\,2$p$ signal does not exceed the noise level. The oxigen peak  (Fig.\,\ref{fig:XPS-HR}\,d) appears  mainly due to surface contamination (e.g., because of the presence of adsorbates); however, a component of the  O\,1$s$ peak with $E_{\mathrm{b}}=531.9$\,eV, which corresponds to the binding energy of electrons in PtO$_2$, is present in the spectrum and the area under it is equal to a  half of the area under the Pt\,4$f$ peak of platinum dioxide. In addition, a component which might be associated with silicon dioxide may also be discriminated in the spectrum. Note, that components of O\,1$s$ peak  attributed PtO$_2$ and SiO$_2$ are overlapped with components which may be associated with surface contaminants that introduces some uncertainty in interpretation of data relating to this peak (Fig.\,\ref{fig:XPS-HR}\,d,\,e). 

The C\,1$s$ signal (Fig.\,\ref{fig:XPS}, curve\,1) is likely completely determined by surface contaminants adsorbed during the sample transportation through the air into the analytical chamber.

Annealing of the Pt/poly-Si structure for formation of the Pt polycide considerably transforms the survey spectrum (Fig.\,\ref{fig:XPS}, curve 2). At first, Si\,2$p$ and Si\,2$s$ peaks appear in the spectrum. At second, the  O\,1$s$ signal significantly increases, with the absolute magnitude of the Pt\,4$f$ peak noticeably decreasing. Components with maxima at $E_{\mathrm{b}}= 71.1$, 72.6 and 75.1\,eV attributed to Pt, PtSi (and likely Pt$_2$Si, i.e. the peaks corresponding to these silicides cannot be resolved using non-monochromatic X-rays) and Pt oxide are present in the spectrum (Fig.\,\ref{fig:XPS-HR}\,b); the contents of these substances in the film correlate as $86:11:3$. Thus, the fraction of Pt silicide grows more than twice whereas the fraction of Pt dioxide decreases by about two times as a result of this technological operation.

The analysis of the Si\,2$p$ peak demonstrates that it can be deconvolved into two components with maxima at $E_{\mathrm{b}}= 102.5$ and 103.4\,eV (Fig.\,\ref{fig:XPS-HR}\,c); the more intense component may be attributed to the natural oxide. This assumption is proved by the presence of a component corresponding to SiO$_2$ in the O\,1$s$ peak (Fig.\,\ref{fig:XPS-HR}\,e); its intensity  slightly exceeds the value which might be explained by the presence of SiO$_2$ that might be attributed, e.g., to the presence of the surface adsorbate for which oxygen peaks have the same chemical shifts as in SiO$_2$. The less intense component of the Si\,2$p$ peak (Fig.\,\ref{fig:XPS-HR}\,c) corresponds to the Si$^{3+}$ state. The concentration of the Si atoms in this state is close to the concentration of the Pt atoms in platinum silicide that allows us to assume the formation of more complex compounds than PtSi and Pt$_2$Si in the topmost layer of the Pt silicide/poly-Si structure such as, e.g., Pt silicates\cite{Pt-silicate_Optoelectronic_Sensors} or some other amorphous substances composed by Pt, Si and O which cannot be detected by the X-ray diffractometry.

The Mo peaks in the discussed spectra in Figs.\,\ref{fig:XPS} and \ref{fig:XPS-HR}\,d,\,e relate to the sample holder. 

Taking into the account the presence of intense peaks of Pt, Si and O we can conclude from the XPS analysis of the Pt silicide/poly-Si structure that its topmost layer is formed by a non-uniform film composed by Pt, SiO$_2$ and some compounds of Pt, Si and O having a complex structure. A thickness of this film exceeds the  depth of sensitivity of XPS that is about 10\,nm. This conclusion well agrees with the observations and analysis made above by means of STEM and X-ray reflectometry. We should emphasize also that SiO$_2$, Pt silicates, etc. may be amorphous and therefore they are not registered by  the X-ray diffractometry.

\begin{figure}[t]
\includegraphics[scale=.97]{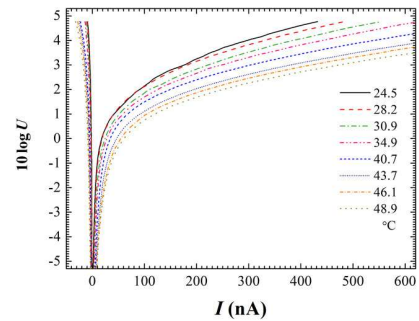}
\caption{\label{fig:I-V}(Color online) 
 Logarithmic \textit{V--I} characteristics of the PtSi/poly-Si diode measured at different temperatures; the voltage $U$ is measured in volts; the temperature in degrees Celsius corresponding to each curve is indicated in the plot.
}
\end{figure}

\begin{figure}[t]
\includegraphics[scale=.95]{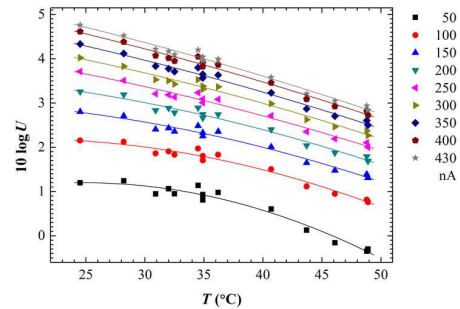}
\caption{\label{fig:logU}(Color online) 
Temperature dependences of voltages for fixed forward currents through the PtSi/poly-Si Schottky diode in the interval from 25 to 50\,{\textcelsius}; the voltage $U$ under the logarithm is measured in volts; the legend to the right of the graph represents the amperage for each curve.
}
\end{figure}

\subsection{R\'{e}sum\'{e} of Analysis}

We can conclude now from the above analysis that (i) the initial Pt/poly-Si film contains a Pt$_2$Si sublayer, about 6\,nm thick,  which forms at room temperature at the interface of Pt and poly-Si  during Pt deposition on poly-Si by magnetron sputtering.
(ii) About 15\,nm thick granular film has formed on top of the Pt silicide as a result of the polycide formation in the solid-state reaction of the thin film of Pt with poly-Si; it is composed by grains of Pt, Si, Pt silicides and amorphous compounds of Pt, Si and O. 
(iii) The PtSi phase dominates in the composition of the resultant Pt silicide and  electrical properties of the produced Schottky diode should be determined by the PtSi/poly-Si junction.
The obtained structure will be further referred to as PtSi/poly-Si.

\section{Electrical Properties and Temperature Sensitivity}

\subsection{\textit{I--V} Characteristics at Different Temperatures}

$I$--$V$ characteristics of the PtSi/poly-Si Schottky diodes (Fig.\,\ref{fig:process}\,k,\,l) measured at different temperatures in the temperature interval 24 to 50{\,\textcelsius} are shown in Fig.\,\ref{fig:I-V} as logarithmic \textit{V--I} curves for several values of the temperature. The curves in Fig.\,\ref{fig:I-V} were obtained from one diode randomly selected from a large number of produced chips of diodes of different shapes and sizes; all the tested diodes had similar \textit{I--V} characteristics, but some parameters varied from chip to chip, e.g., the rectification ratios varied from several hundreds to more than a thousand for different diode sizes. However, in general the dependances of diode current on temperature were similar and the values characterizing the temperature sensitivity, such as \textit{TCU}, were close for the examined chips. So, since we do not set a goal to investigate the production statistics, such as potential yield of the diodes, we will demonstrate the main parameters of the PtSi/poly-Si Schottky diodes  analysing an example of only one but representative sample.

\begin{figure}[t]
\includegraphics[scale=1]{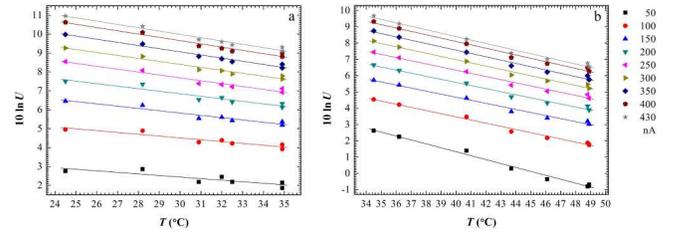}
\caption{\label{fig:lnU-fitting}(Color online) 
Linear fits of dependences of voltage logarithms on the PtSi/poly-Si Schottky diode temperature for fixed forward currents through the diode in two temperature intervals: (a) close to the room temperature (from 25 to 35\,{\textcelsius}) and (b) at elevated temperatures (from 35 to 50\,{\textcelsius}); the voltage $U$ under $\ln$ is taken in volts; the legend to the right of the plots represents current values corresponding to the fitted sets of the experimental points.
}
\end{figure}

The \textit{V--I} curves measured at different temperatures allowed us to derive dependances of the voltage drop across the diode on temperature for different values of current flowing through the diode (Fig.\,\ref{fig:logU}). The obtained  $\log U(T)$ curves for the forward biased diode are seen to be non-linear for the temperatures from 25 to 50{\,\textcelsius} and current values up to at least 400\,nA. However, we can see a point around 35\,{\textcelsius} which divides the curves into two nearly linear parts; if we subdivide the temperature interval into two narrower ones at this point we obtain two sets of straight lines for each of these intervals  which enable us to obtain the values of \textit{TCU} (Fig.\,\ref{fig:lnU-fitting}).

\begin{figure}[t]
\includegraphics[scale=.95]{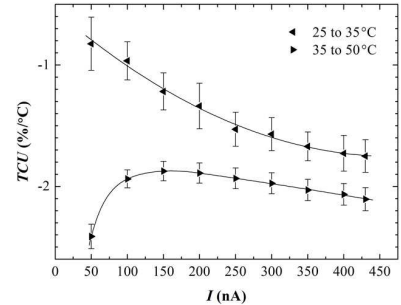}
\caption{\label{fig:TSU}
Temperature coefficients of voltage $(TCU)$ derived from linear fitting of temperature dependences of voltage drop on the PtSi/poly-Si Schottky diode in the temperature intervals from 25 to 35 and from 35 to 50\,{\textcelsius} (Fig.\,\ref{fig:lnU-fitting}) represented as functions of the forward current through the diode.
}
\end{figure}

\subsection{Temperature Coefficient of Voltage}

The values of \textit{TCU} for different currents through the diode are calculated by linear fitting the data of the $\ln U(T)$ dependances (Fig.\,\ref{fig:lnU-fitting}). The slopes of the obtained straight lines determine the functions $TCU(I)=d[\ln U(I,T)]/dT$ for each of the temperature intervals. These curves are presented in Fig.\,\ref{fig:TSU}. The values of \textit{TCU} are seen to be negative for all the values of current in the both temperature intervales.

In the temperature interval from 25 to 35\,{\textcelsius}, the absolute value of \textit{TCU} initially non-linearly grows with increasing current until saturates at nearly 2\,\%/{\textcelsius} when the current is increased to about 400\,nA. If the temperature is in the interval from 35 to 50\,{\textcelsius}, initially high in absolute values \textit{TCU} decreases, reaches minimum at about 150\,nA and then increases again reaching the absolute values some greater than 2\,\%/{\textcelsius}. So, we can conclude that there exist an interval of current in which  \textit{TCU} is nearly permanent in  the whole temperature interval from 25 to 50\,{\textcelsius} in which we have investigated the PtSi/poly-Si diode electrical properties.\footnote{This behavior of \textit{TCU}  with $I$ and $T$ may be explained by changes in the states at the grain boundaries of the polycrystalline Schottky diodes.\cite{Poly_Semiconductors-De_Graaff}}

The obtained values of \textit{TCU} allow us to conclude that the PtSi/poly-Si Schottky diodes are very promising temperature sensors for bolometers having sufficiently high temperature sensitivity.  \textit{TCU} of a bolometer cell, and hence its sensitivity to a detected radiation,  can be easily increased by placing a set of serially connected diodes on a membrane.\footnote{\textit{TCR\,=\,TCU} is not the only figure of merit for bolometers. Signal-to-noise ratio is important as well. We predict a relatively high $1/f$-noise in these bolometers like it takes place in polycrystalline VO$_x$ or SiGe bolometers.\cite{Akin_CMOS_Thermal} At the same time we expect an increase in signal-to-noise ratio due to size reduction  and a large number of diodes on membranes.}

\section{Conclusion}
Summarizing the above we should emphasize the main results of the article.

First, using a technological process well developed in the industry, which is compatible with the CMOS or CCD manufacturing process, we have formed thin-film PtSi/poly-Si (polycide) Schottky junctions on artificial dielectric Si$_3$N$_4$/SiO$_2$/Si(001):B substrates simulating the bolometer membranes and ready to MEMS-etching the bolometer cells. 

Second, the analysis of the polycide formation process has demonstrated that the initial Pt/poly-Si film contains a Pt$_2$Si sublayer, about 6\,nm thick,  which forms at room temperature at the interface of Pt and poly-Si  during Pt deposition by magnetron sputtering. 
In the process of the Pt polycide formation in thin films, a granular film forms on top of the Pt silicide (in this work, it was 15\,nm thick); it turned out to be composed by grains of Pt, Si, Pt silicides and amorphous compounds of Pt, Si and O.
The PtSi phase dominates in the resultant Pt silicide and  electrical properties of the produced Schottky diode are determined by the PtSi/poly-Si junction.
  
And finally, we have demonstrated that the obtained PtSi/poly-Si  Schottky diodes have sufficient temperature sensitivity ($TCU\approx 2$\,\%/{\textcelsius} at optimal operating conditions) to be used as  temperature sensors in  bolometer detectors of radiation. Their sensitivity exceeds that of standard \textit{p--n}-junctions formed in single-crystalline Si or in SOI substrates whereas their manufacturability in  MEMS processes used for bolometer fabrication is higher than that of mono-Si diodes.

\begin{acknowledgments}
We thank Prof. A. G. Touryanski for his support of the X-ray analyses and  Dr. V. V. Voronov for fruitful discussion the X-ray data.
This research is  a part of the collaborative research project funded by the RF Government (contract No.\,63/CM/2012).
We thank Ms. N. V. Kiryanova, TP GPI CEO and Deputy Head of the IR Technology Dept. of GPI RAS, for her contribution to management of this research. 
The research was in part supported by the Marie Sk{\l}odowska-Curie International Incoming Fellowship within the 7th European Community Framework Programme (call ref. FP7-PEOPLE-2011-IIF, 
project No. 298932);
Peter Gaiduk 
thanks for funding.
The equipment of the Center for Collective Use of Scientific Equipment of GPI RAS was used for this study.


\end{acknowledgments}



%




\end{document}